\documentclass[11pt, journal, onecolumn]{IEEEtran}

\usepackage{fullpage}
\usepackage{setspace}
\usepackage{amsmath,amssymb,mathptm}
\usepackage{times}
\usepackage{latexsym}
\usepackage{url}
\usepackage{graphicx}
\usepackage{color}

\input{epsf}
\input{isolatin1.sty}

\newtheorem{theorem}{Theorem}
\newtheorem{definition}{Definition}
\newtheorem{corollary}{Corollary}

\def\Reals{I\!\!R}
\newcommand{\bvec}[1]{\boldsymbol{#1}}
\newcommand{\E}[1]{\mathbb{E}\!\left(#1\right)}
\newcommand{\pr}[1]{\mathbb{P}\!\left(#1\right)}

\begin{document}

\title{Dynamic Coverage of Mobile Sensor Networks}

\author{Benyuan~Liu, Olivier~Dousse, Philippe~Nain and~Don~Towsley
\thanks{Benyuan Liu is with the Department of Computer Science, University of Massachusetts Lowell, email: bliu@cs.uml.edu. Olivier Dousse is with Nokia Research Center, Lausanne, Switzerland, email: odousse@mac.com.  Philippe Nain is with INRIA Sophia Antipolis, France, email: Philippe.Nain@sophia.inria.fr. Don Towsley is with the Department of Computer Science, University of Massachusetts, Amherst, email: towsley@cs.umass.edu.}}

\maketitle

\begin{abstract}
In this paper we study the dynamic aspects of the coverage of a mobile sensor 
network resulting from continuous movement of sensors. As sensors move around, 
initially uncovered locations are likely to be covered at a later time. A 
larger area is covered  as time continues, and intruders that might 
never be detected in a stationary sensor network can now be detected
by moving sensors. However, this improvement in coverage is achieved at the 
cost that a location is covered only part of the time, alternating between covered and not 
covered. We characterize area coverage at specific time instants 
and during time intervals, as well as the time durations that a location
is covered and uncovered. We further characterize the time it takes to 
detect a randomly located intruder. For mobile intruders, we take a game 
theoretic approach and derive optimal mobility strategies for both 
sensors and intruders. Our results show that sensor mobility brings about
unique dynamic coverage properties not present in a stationary sensor 
network, and that mobility can be exploited to compensate for the lack of 
sensors to improve coverage.
\end{abstract}


\section{Introduction}

Recently, there has been substantial research in the area of sensor 
network coverage. The coverage of a sensor network represents the 
quality of surveillance that the network can provide, for example, 
how well a region of interest is monitored by sensors, and how 
effectively a sensor network can detect intruders. It is important 
to understand how the coverage of a sensor network depends on various 
network parameters in order to better design and use sensor networks 
for different application scenarios.

In many applications, sensors are not mobile and remain stationary
after their initial deployment. The coverage of such a stationary sensor 
network is determined by the initial network configuration. Once the 
deployment strategy and sensing characteristics of the sensors are known, 
 network coverage can be computed and remains unchanged over time.
 
Recently, there has been increasing interest on building mobile sensor networks.
Potential applications abound. Sensors can be mounted on mobile platforms 
such as mobile robots and move to desired areas \cite{Serment02Millibots,Bergbreiter03CotsBots,Gabriel02Robomote,Matalin04Coverage}. Such mobile 
sensor networks 
are extremely valuable in situations where traditional deployment mechanisms 
fail or are not suitable, for example, a hostile environment where sensors 
cannot be manually deployed or air-dropped. 
Mobile sensor networks can also play a vital role in homeland security. Sensors 
can be mounted on vehicles (e.g., subway trains, taxis, police cars, fire 
trucks, boats, etc) or carried by people (e.g., policemen, fire 
fighters, etc). These sensors will move with their carriers, dynamically 
patrolling and monitoring the environment (e.g., chemical, biological, or 
radiological agents). In other application scenarios such as atmosphere and 
under-water environment monitoring, airborne or under-water sensors may 
move with the surrounding air or water currents \cite{Sanderson05RiverNet}. 
The coverage 
of a mobile sensor network now depends not only on the initial network 
configurations, but also on the mobility behavior of the sensors.

While the coverage of a sensor network with stationary sensors has been 
extensively explored and is relatively well understood, researchers have only 
recently started to study the coverage of mobile sensor networks. Most of 
this work focuses on algorithms to reposition sensors in desired positions 
in order to enhance network coverage \cite{howard02mobile,Batalin02Spreading,Pearce03Dispersion,Zou03Sensor,Wang04Movement}. More specifically, 
these proposed algorithms strive to spread sensors to desired locations to 
improve coverage. The main differences 
among these works are how exactly the desired positions of sensors are 
computed. Although the algorithms can adapt to changing environments and 
recompute the sensor locations accordingly,  sensor mobility is exploited
essentially to obtain a new stationary configuration that improves
coverage after the sensors move to their desired locations. 

In this paper, we study the coverage of a mobile sensor network from a
different perspective. Instead of trying to achieve an improved stationary 
network configuration as the end result of sensor movement, we are interested
in the dynamic aspects of network coverage resulting from the continuous 
movement of sensors. 
In a stationary sensor network, the covered areas are determined by the 
initial configuration and do not change over time. In a mobile sensor 
network, previously uncovered areas become covered as sensors move 
through them and covered areas become uncovered as sensors move away. 
As a result, the areas covered by sensors change over time, and more 
areas will be covered at least once as time continues. The coverage 
status of a location also changes with time, alternating between being covered 
and not being covered. In this work, we assume that sensors are initially randomly 
and uniformly deployed and move independently in randomly chosen 
directions. Based on this model, we characterize the fraction of area 
covered at a given time instant, the fraction of area ever covered 
during a time interval, as well as the time durations that a location 
is covered and not covered. 

Intrusion detection is an important task in many sensor network 
applications. We measure the intrusion detection capability of a mobile
sensor network by the detection time of a randomly located intruder, 
which is defined to be the 
time elapsed before the intruder is first detected by a sensor. In a stationary 
sensor network, an initially undetected intruder will 
never be detected if it remains stationary or moves along an uncovered 
path. In a mobile sensor network, however, such an intruder may be 
detected as the mobile sensors patrol the field. This can significantly 
improve the intrusion detection capability of a sensor network. In this paper we 
characterize the detection time of a randomly located intruder. The 
results suggest that sensor mobility can be exploited to effectively 
reduce the detection time of a stationary intruder when the number of 
sensors is limited. We further present a lower bound on the distribution
of the detection time of a randomly located intruder, and show that it can be 
minimized if sensors move in straight lines. 

In some applications, for example, radiation, chemical, and biology agents
detections, there is a sensing time requirement before an intruder is detected. 
We find in this case that too much mobility can be harmful if the sensor speed is above
a threshold. Intuitively, if a sensor moves faster, it will cover an 
area more quickly and detect some intruders sooner, however, at the same time, 
it will miss some intruders due to the sensing time requirement. To this end,
we find there is an optimal sensor speed that minimizes the detection time of 
a randomly located intruder.

For a mobile intruder, the detection time depends on the mobility strategies of
both  sensors and intruder. We take a game theoretic approach and 
study the optimal mobility strategies of sensors and intruder. Given the 
sensor mobility pattern, we assume that an intruder can choose its mobility
strategy so as to maximize its detection time (its lifetime before being 
detected). On the other hand, sensors choose a mobility strategy that 
minimizes the maximum detection time resulting from the intruder's mobility 
strategy. This can be viewed as a zero-sum minimax game between the collection of mobile sensors 
and the intruder. We prove that the optimal sensor mobility strategy is for
sensors to choose their directions uniformly at random between $[0, 2\pi)$.   The 
corresponding intruder mobility strategy is to remain stationary to maximize 
its detection time. This solution represents a mixed strategy which is a Nash 
equilibrium of the game between mobile sensors and intruders. If sensors 
choose to move in any fixed direction (a pure strategy), it can be exploited by an 
intruder by moving in the same direction as sensors to maximize its detection time. 
The optimal sensor strategy is to choose a mixture of available pure 
strategies (move in a fixed direction between $[0, 2\pi)$). The proportion 
of the mix should be such that the intruder cannot exploit the choice by 
pursuing any particular pure strategy (move in the same direction as 
sensors), resulting in a uniformly random distribution for sensor's movement. 
When sensors and intruders follow their respective optimal strategies, 
neither side can achieve better performance by deviating from this behavior.

The remainder of the paper is structured as follows. In Section \ref{sec:sensor_relatedwork}, 
we review related work on the coverage of sensor networks. The network model  
and coverage measures are defined in Section \ref{sec:model}. In 
Section \ref{sec:areaCoverage}, we derive the fraction of the area being 
covered at specific time instants and during a time interval. 
The detection time for both stationary and mobile intruders are 
studied in Section \ref{sec:StationaryDetectionTime} and Section 
\ref{sec:MobileDetectionTime}, respectively. In Section 
\ref{sec:MobileDetectionTime}, we also derive the the optimal mobility
strategies for sensors and intruders from a game theoretic perspective.
 Finally, we summarize the paper in Section 
\ref{sec:conclusions}. 

\section{Related Work} \label{sec:sensor_relatedwork}

Recently, sensor deployment and coverage related topics have become an
active research area. In this section, we present a brief overview of
the previous work on the coverage of both stationary and mobile sensor
networks that is most relevant to our study. A more thorough survey of 
the sensor network coverage problems can be found in \cite{Cardei04Coverage}.

Many previous studies have focused on characterizing various coverage
measures for stationary sensor networks. In \cite{Shakkottai03sensor}, 
the authors considered a grid-based sensor network and derived the 
conditions for the sensing range and failure rate of sensors to ensure 
that an area is fully covered. In \cite{meguerdichian01coverage}, the 
authors proposed several algorithms to find paths that are most or least 
likely to be detected by sensors in a sensor network. Path exposure of 
moving objects in sensor networks was formally defined and studied in 
\cite{meguerdichian01exposure}, where the authors proposed an
algorithm to find minimum exposure paths, along which the probability
of a moving object being detected is minimized. The path exposure problem
is further explored in \cite{Li03Coverage,Veltri03Minimal,Chin05Exposure}. 
In \cite{Huang03Coverage,Zhou04Connected,Kumar04kcoverage}, the 
$k$-coverage problem where each point is covered by at least $k$ sensors 
was investigated. In \cite{Liu04Coverage}, the authors defined and derived
several important coverage measures for a large-scale stationary sensor 
network, namely, area coverage, detection coverage, and node coverage,
under a Boolean sensing model and a general sensing model. Other coverage 
measures have also been studied. In \cite{Gui04power,Gui05Virtual}, the 
authors studied a metric of quality of surveillance which is defined to be 
the average distance that an intruder can move before being detected,
and proposed a virtual patrol model for surveillance operations in sensor
networks. The relationship between area coverage and network connectivity is 
investigated in \cite{Wang03Integrated, Zhang04Maintaining, Bai06Deploying}.

While the coverage of stationary sensor networks has been extensively
studied and relatively well understood, researchers have started to explore
the coverage of mobile sensor networks only recently. In 
\cite{howard02mobile,Zou03Sensor,Poduri04Constrained}, 
virtual-force based algorithms are used to repel nodes from each other and 
obstacles to maximize coverage area. In \cite{Wang04Movement}, algorithms 
are proposed to identify existing coverage holes in the network and compute 
the desired target positions where sensors should move in order to increase 
the coverage. In \cite{Cortes04Coverage}, a distributed control and coordination 
algorithm is proposed to compute the optimal sensor deployment for a class of 
utility functions which encode optimal coverage and sensing policies. In 
\cite{Zhang05Controlling}, mobility is used for sensor density control such 
that the resultant densor density follows the spatial variation of a scalar 
field in the environment. In \cite{Tang04Planning}, an autonomous planning 
process is developed to compute the deployment positions of sensors and leader 
waypoints for navigationally-chanllenged sensor nodes. The deployment of 
wireless sensor networks under mobility constraints and the tradeoff between 
mobility and sensor density for coverage are studied in \cite{Chellappan07Deploying,Wang07Tradeoff}. 

Many of these proposed algorithms strive to spread sensors to desired 
positions in order to obtain a stationary configuration such that the 
coverage is optimized. The main difference is how the desired sensor
positions are computed. In this work we study the coverage of a mobile 
sensor network from a very different perspective. Instead of trying to 
achieve an improved stationary network configuration as an end result 
of sensor movement, we focus on the dynamic coverage properties resulting 
from the continuous movement of the sensors. 

Intrusion detection problem in mobile sensor networks has been considered 
in a few recent studies, e.g., \cite{Liu05Coverage, Chin06Analytic, Dousse06Delay, 
Keung10Intrusion, Peres10Mobile}.
In our work we take a stochastic geometry based approach to derive closed-form 
expressions for the detection time under different network, mobility, and sensing 
models. In \cite{Chin10Detection}, Chin et. al. proposed and studied a similar game 
theoretic problem formulation for a different network and mobility model.

\section{Network and Mobility Models} \label{sec:model}

In this section, we describe the network and mobility model, and 
introduce three coverage measures for a mobile sensor network used 
in this study. 

\subsection{Sensing Model}

We assume that each sensor has a sensing radius, $r$. A sensor can only 
sense the environment and detect intruders within its sensing area, which 
is the disk of radius $r$ centered at the sensor. A point is said to be 
{\em covered} by a sensor if it is located in the sensing area of the 
sensor. The sensor network is thus partitioned into two regions, the covered 
region, which is the region covered by at least one sensor, and the 
uncovered region, which is the complement of the covered region. An 
intruder is said to be {\em detected} if it lies within the covered 
region.

In reality, the sensing area of a sensor is usually not of disk shape due
to hardware and environment factors. Nevertheless, the disk model can be
used to approximate the real sensing area and provide bounds for the real
case. For example, the irregular sensing area of a sensor can be lower and 
upper bounded by its maximum inscribed and minimum circumscribed circles, 
repectively.

\subsection{Location and Mobility Model}

We consider a sensor network consisting of a large number of sensors 
placed in a 2-dimensional infinite plane. This is used to model a large 
two-dimensional geographical region. For the initial 
configuration, we assume that, at time $t = 0$, the locations of these 
sensors are uniformly and independently distributed in the region. Such 
a random initial deployment is desirable in scenarios where prior knowledge 
of the region of interest is not available; it can also result from  certain 
deployment strategies.  Under this assumption,
the sensor locations can be modeled by a stationary two-dimensional Poisson 
point process. Denote the density of the underlying Poisson point process 
as $\lambda$. The number of sensors located in a region $R$, $N(R)$, follows 
a Poisson distribution with parameter $\lambda \|R\|$, where $\|R\|$ 
represents the area of the region.  

Since each sensor covers a disk of radius $r$, the initial
configuration of the sensor network can be described by a Poisson Boolean
model $B(\lambda, r)$. In a stationary sensor network, sensors do not move
after being deployed and network coverage remains the same as that of
the initial configuration. In a mobile sensor network, depending on the
mobile platform and application scenario, sensors can choose from a wide
variety of mobility strategies, from passive movement to highly coordinated
and complicated motion. For example, sensors deployed in the air or water
may move passively according to external forces such as air or water
currents; simple robots may have a limited set of mobility patterns, and
advanced robots can navigate in more complicated fashions; sensors mounted
on vehicles and people move with their carriers, which may move randomly and 
independently or perform highly coordinated search.
 
In this work, we consider the following sensor mobility model. Sensors follow 
arbitrary random curves independently of each other without coordination 
among themselves. In some cases, when it helps to yield closed-form results 
and provide insights, we will make the model more specific by limiting sensor
movement to straight lines. In this model, the movement of a sensor is 
characterized by its speed and direction. A sensor randomly chooses a 
direction $\Theta \in [0, 2\pi)$ according to some distribution with a 
probability density function of $f_{\Theta}(\theta)$. The speed of the 
sensor, $V_s$, is randomly chosen from a finite range  $[0, v_s^{\max}]$, 
according to a distribution density function of $f_{V_s}(v)$. The sensor speed 
and direction are independently chosen from their respective distributions.

The above models make simplified assumptions for real network scenarios. 
Our purpose is to obtain analytical results based on the simplied assumptions 
and provide insight and guideline to the deployment and performance of mobile 
sensor networks. The Poisson distribution and unit disk model have been widely 
used in the studies of wireless networks (e.g., coverage and capacity problems)
to obtain analytical results. The Poisson spatial distribution is a good 
approximation for large networks where nodes are randomly and uniformly 
distributed. For the mobility model, we consider the scenarios where 
nodes move independently of each other. For example, sensors can be carried 
by people or mounted on people's vehicles, boats, or animals, etc.
These carriers are likely to move independently according to their own 
activity patterns without much coordination. This is similar to the 
{\it uncoordinated mobility model} used in \cite{Chin06Analytic}. Note that
in some scenarios (e.g., sensors mounted on robots) mobile sensors can 
communicate with each other and coordinate their moves. In that case the 
sensors can optimize their movement patterns and provide more efficient 
coverage than the independent mobility case. In this paper we will focus 
on the independent mobility model. 

Throughout the rest of this paper, we will refer to the initial
sensor network configuration as {\it random sensor network $B(\lambda, r)$}, 
the first mobility model where sensors move in arbitrary curves as 
{\it random mobility model}, and the more specific mobility model where 
sensors all move in straight lines as {\it straight-line mobility 
model}.  The shorthand $X \sim \exp(\mu)$ stands for $\pr{X < x}=1 - \exp(-\mu x)$, i.e., random variable $X$ is 
exponentially distributed with parameter $\mu$.

\subsection{Coverage measures}

To study the dynamic coverage properties of a mobile sensor network, we
define the following three coverage measures. 

\begin{definition}
{\bf Area coverage}: The area coverage of a sensor network at time $t$, $f_a(t)$, is the probability that a given point $\bvec{x} \in \Reals^2$ is covered by one or more sensors at time $t$.
\end{definition}

\begin{definition}
{\bf  Time interval area coverage}: The area coverage of a sensor network during time interval $[s,t)$ with $s<t$, $f_i(s,t)$, is the probability that given a point $x\in\Reals^2$, there exists $u\in[s,t)$ such that $x$ is covered by at least one sensor at time $u$.
\end{definition}

\begin{definition}
{\bf Detection time}: Suppose that an intruder has a trajectory $x(t)$ and that $x(0)$ is uncovered at time $t=0$. The detection time of the intruder is the smallest $t>0$ such that $x(t)$ is covered by at least one sensor at time $t$.
\end{definition}

All three coverage measures depend not only on static
properties of the sensor network (initial sensor distribution, sensor density and sensing range),
but also on sensor  movements. The characterization of
area coverage at specific time instants is important for applications
that require parts of the whole network be covered at any given time instant. The time interval area coverage 
 is relevant for applications that do not require 
or cannot afford simultaneous coverage of all locations 
at specific time instants, but prefer to cover the network within some 
time interval. The detection time is important for intrusion detection
applications, measuring how quickly a sensor network can detect a 
randomly located intruder.

\section{Area Coverage} \label{sec:areaCoverage}

In this section, we study and compare the area coverages of both stationary
and mobile sensor networks. We first analytically characterize  the area 
coverage. We then discuss the implications of our results on network planning 
and show 
that sensor mobility can be exploited to compensate for the lack of sensors 
to increase the area being covered during a time interval. However, we point 
out, due to the sensor mobility, a point is only covered part of the time; 
we further characterize this effect by determining the fraction of time that 
a point is covered. Finally, we discuss the optimal moving strategies that 
maximize the area coverage during a time interval.

\begin{figure} 
  \begin{center}
    \includegraphics[width=3.6in]{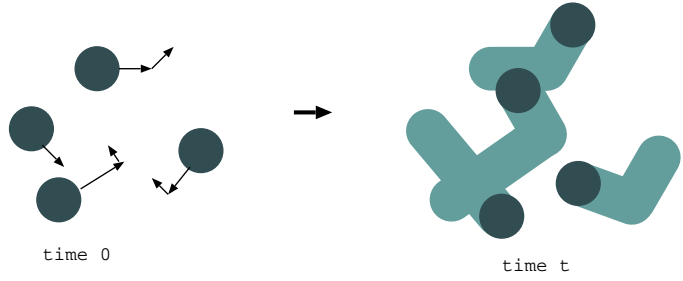}
    \caption{Coverage of mobile sensor network: the left figure depicts the
      initial network configuration at time 0 and the right figure
      illustrates the effect of sensor mobility during time interval
      $[0,t)$. The solid disks constitutes the area being covered at
      the given time instant, and the union of the shaded region and
      the solid disks represents the area being covered during the time
      interval.}
    \label{fig_mobile_sensors}
  \end{center}
\end{figure}

In a stationary sensor network, a location always remains either covered or
not covered. The area coverage does not change over time. The effect of 
sensor mobility on area coverage is illustrated in Figure 
\ref{fig_mobile_sensors}. The union of the solid disks constitutes the 
area coverage at given time instants. As sensors move around, exact 
locations that are covered at different time instants change over time. 
The area that has been covered during time interval $[0,t)$ is 
depicted as the union of the shaded region and the solid disks. As can 
be observed, more area is covered during the time interval than the 
initial covered area. The following theorem characterizes the effect of 
sensor mobility on area coverage.

{\bf
\begin{theorem} \label{theo:msc}
Consider a sensor network $B(\lambda, r)$ at time $t=0$,
with sensors moving according to the random mobility model. 
\begin{enumerate}
\item  At any time instant $t$, the fraction of area being covered is
\begin{eqnarray}\label{eq:fat}
f_a(t) = 1 - e^{-\lambda \pi r^2},\;\; \forall t\geq 0.
\end{eqnarray}
\item The fraction of area that has been covered at least once during time
interval $[s,t)$ is 
\begin{eqnarray}\label{eq:fit}
f_i(s,t) = 1 - e^{-\lambda \E{\alpha(s,t)}}.
\end{eqnarray}
where $\E{\alpha(s,t)}$ is the expected area covered by a sensor during time 
interval $[s,t)$. When sensors all move in straight lines, we have
\begin{eqnarray}\label{eq:fit2}
f_i (s,t) = 1 - e^{-\lambda (\pi r^2 + 2r\bar{v}_s(t-s))}.
\end{eqnarray}
where  $\bar{v}_s$ is the average sensor speed.
\item The fraction of the time a point is covered is 
\begin{eqnarray} \label{eq:ft}
f_t = 1-e^{-\lambda \pi r^2}.
\end{eqnarray}
\end{enumerate}
\end{theorem}
}

{\bf Proof.} 
Denote the initial sensors location process as $N_0$.  By assumption, $N_0$ is a two-dimensional Poisson point process of density $\lambda$ over $\Reals^2$.
  The probability that a point $\bvec{x}\in \Reals^2$ is covered is equal to the probability that at least one sensor is located in the disk of radius $r$ centered on $\bvec{x}$ (denoted hereafter $D(\bvec{x},r)$).
  At time $t=0$, the location of the sensors is given by $N_0$, and we have $f_a(0) = \pr{N_0 \cap D(\bvec{x},r) \neq \emptyset} = 1-\exp(-\lambda \pi r^2)$.

As illustrated in Figure \ref{fig_mobile_sensors}, during a time interval, 
each sensor covers a greater area than its sensing area at specific time 
instants. Denote the covered area during time interval $[s,t)$ by 
$\alpha(s,t)$. The area coverage during the time interval can be computed as
\begin{eqnarray*}
f_i(s,t) = 1 - e^{-\lambda \E{\alpha(s,t)}}.
\end{eqnarray*}

If sensors all move in straight lines, during time interval $[s,t)$, each 
sensor has covered a shape of a racetrack whose expected area is 
\begin{eqnarray*}
\E{\alpha(s,t)} = \E{\pi r^2+2rv_s(t-s)}=\pi r^2 + 2r\bar{v}_s(t-s).
\end{eqnarray*}
where $\bar{v}_s = \int_0^{v_s^{\max}}f_V(V) dV$ represents the expected
sensor speed. In this case, the area coverage is given by
\begin{eqnarray*}
f_i(s,t)= 1 - e^{-\lambda (\pi r^2 +2r \bar{v}_s(t-s))}. 
\end{eqnarray*}

While an uncovered location will be covered when a sensor moves within
distance $r$ of the location, a covered location  becomes
uncovered as  sensors covering it move away. As a result, a location
is only covered part of the  
time. More specifically, a location alternates between  being 
covered and not being covered, which can be modeled as an alternating 
renewal process.  We use the fraction of time that a location is covered 
to measure this  effect. As $f_a(t)=1-\exp(-\lambda \pi r^2)$ for 
all $t$, the expected fraction of time that a location is covered, 
$f_t$, is also equal to $1-\exp(-\lambda \pi r^2)$.

The case where sensors and intruder are confined in a finite area can be 
addressed by considering a finite sample of the model presented in this 
paper. The average fraction of the area being covered at time $t$ (or 
respectively during the interval $[s,t]$) is equal to $f_a(t)$ ($f_i(s,t)$ 
respectively). Furthermore, as the sensor location process at time $t$, $N_t$, 
is an ergodic point process, if the sample area grows to infinity, the 
variance of the covered fraction tends to zero. This means that if the 
network area tends to infinity, the covered fraction becomes deterministic.

\hfill $\Box$

Note that this theorem is a generalization of Theorem 2 in \cite{Liu05Coverage}.
At any specific time instant, the fraction of the area being covered by the
mobile sensor network described above is the same as in a stationary
sensor network. This is because at any time instant, the positions of
the sensors still form a Poisson point process with the same parameters as 
in the initial configuration. However, unlike in a stationary sensor 
network, covered locations change over time; areas initially not covered 
will be covered as sensors move around. Consequently, intruders in the 
initially uncovered areas can be detected by the moving sensors.

When sensors all move in straight lines, the fraction of the area that has 
ever been covered increases and approaches one as time proceeds. Later in 
this section we will prove that, among all possible curves, straight line 
movement is an optimal strategy that maximizes the area being covered during 
a time interval. The rate at which the covered area increases over time depends on 
the expected sensor speed. The faster sensors move, the more
quickly the deployed region is covered. Therefore, sensor mobility can be
exploited to compensate for the lack of sensors to improve the area
coverage over an interval of time. This is useful for applications that 
do not require or cannot afford simultaneous coverage of all locations at 
any given time, but need to cover a region within a given time interval. 
Note that the area coverage during a time interval does not depend on the 
distribution of sensors movement direction. Based on (\ref{eq:fit2}), 
we can compute the expected sensor speed required to have a certain 
fraction of the area ($f_0$) covered within a time interval of length 
$t_0$.  
\begin{eqnarray*} 
\bar{v}_s = -\frac{\lambda \pi r^2 + \log (1-f_0)}{2\lambda r t_0},\;\; \mbox{for $f_0 \geq 1-e^{\lambda \pi r^2}$}.
\end{eqnarray*}

However, the benefit of a greater area being covered at least once during 
a time interval comes with a price. In a stationary sensor network, a 
location is either always covered or not covered, as determined by its 
initial configuration. In a mobile sensor network, as a result of sensor 
mobility, a location is only covered part of the time, alternating between 
covered and not covered. The fraction of time that a location is covered 
corresponds to the probability that it is covered, as shown in (\ref{eq:ft}). 
Note that this probability is determined by the static properties of the 
network configuration (density and sensing range of the sensors), and does
not depend on sensor mobility. In the next section, we will further 
characterize the duration of the time intervals that a location is 
covered and uncovered.

From the proof of Theorem \ref{theo:msc}, it is easy to see that area
coverage during a time interval is maximized  when sensors move in straight 
lines. This is because, among all possible curves, the area covered by a 
sensor during time interval $[s,t)$, $\alpha(s,t)$, is maximized when the 
sensor moves in a straight line. Based on (\ref{eq:fit}), we have the 
following theorem. 

{\bf
\begin{theorem}
In a sensor network $B(\lambda, r)$ with sensors moving according to the 
random mobility model, the fraction of area covered during any time interval 
$[s,t)$ is maximized when sensors all move in straight lines.
\end{theorem}
}

It is important to  point out that straight line movement is not the only 
optimal strategy that maximizes the area coverage during a time interval. 
There is a family of optimal movement patterns that maximize the coverage. 
We conjecture that the optimal movement patterns have the following 
properties: 1) the local radius of curvature is greater than the sensing 
range $r$ everywhere along the oriented trajectory; 2) if the euclidean 
distance between two points of the curve is less than $2r$, then the distance 
between them along the curve is less than $\pi r$. When these two properties 
are satisfied, the sensing disk of a sensor does not overlap with its 
previously covered areas, and a point will not be covered redundantly by 
the same sensor. The covering efficiency is thus maximized.

\section{Detection Time of Stationary Intruder} \label{sec:StationaryDetectionTime}

The time it takes to detect an intruder is of great importance in many military
and security-related applications. In this section, we study the detection 
time of a randomly located stationary intruder. Detection time for a mobile 
intruder is investigated in the next section. To facilitate the 
analysis and illustrate the effect of sensor mobility on  detection time, 
we consider the scenario where all sensors move at a constant speed 
$v_s$. More general sensor speed distribution scenarios can be approximated 
using the results of this analysis.

We assume that intruders do not initially fall into the coverage area 
of any sensor. Obviously, these intruders will never be detected in a
stationary sensor network. In a mobile sensor network, however, an intruder 
can be detected by sensors passing within a distance $r$ of it, where $r$ is 
the common sensing range of the sensors. The detection time characterizes how 
quickly the mobile sensors can detect a randomly located intruder previously 
not detected. We will first derive the detection time when sensors all move 
in straight lines. We will then consider the case when sensors move according to
arbitrary curves.

{\bf
\begin{theorem} \label{theo:sdt}
Consider a sensor network $B(\lambda,r)$ with sensors moving according to the straight-line random mobility model and a static intruder.
The sequence of times at which a new sensor detects the intruder forms a Poisson process of intensity $2\lambda r \bar{v}_s$, where $\bar{v}_s$ denotes the average sensor speed. As a consequence, the time before first the detection of the intruder is exponentially distributed with the same parameter.
\end{theorem}
}

\begin{figure}
    \begin{center}
      \includegraphics[width=2in]{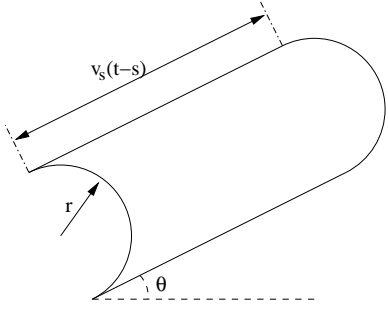}
    \caption{The region $A(s,t)$ under the straight-line mobility model.}
    \label{fig:ast}
    \end{center}
  \end{figure}

{\bf Proof:} We denote by $A(s,t)$ the random region covered by a sensor in the interval $[s,t]$, that was not covered before time $s$.  The shape of this region is illustrated in Figure \ref{fig:ast}.

  We first prove that the number of sensors hitting the intruder in the time interval $[s,t]$ is Poisson distributed with parameter $2\lambda r \bar{v}_s (t-s)$.
  Suppose without loss of generality that the intruder is located at the origin. 
  The probability that a sensor initially located at point $\bvec{x} \in \Reals^2$ hits the intruder within $[s,t]$ is equal to $\pr{-\bvec{x} \in A(s,t)}$. 
  This probability only depends on the direction and speed of the sensors; in
particular, it does not depend on  the initial Poisson process giving the positions of the sensors.
  We can thus define a thinned Poisson process $\Phi(s,t)$ by selecting at time $0$ the sensors that will hit the intruder during the interval $[s,t]$. This process is non-uniform and has density
  $$
  \lambda'(\bvec{x}) = \lambda \pr{-\bvec{x} \in A(s,t)}.
  $$
  The number of sensors hitting the intruder during $[s,t]$ is equal to the total number of points in the thinned process, which is Poisson distributed with mean 
  \begin{eqnarray}
    \E{\text{card}(\Phi(s,t))} &=& \int_{\Reals^2} \lambda'(\bvec{x}) \bvec{dx} \nonumber \\
    &=& \lambda \int_{\Reals^2} \pr{-\bvec{x} \in A(s,t)} \bvec{dx} \nonumber \\
    &=& \lambda \int_{\Reals^2} \E{1_{\{-\bvec{x} \in A(s,t) \} }} \bvec{dx} \nonumber \\
    &=& \lambda \E{ \int_{\Reals^2} 1_{\{-\bvec{x} \in A(s,t) \} } \bvec{dx} } \nonumber \\
    &=& \lambda \E{ || A(s,t) || }, \label{eq:intensity}
  \end{eqnarray}
  where $1_{\{ \cdot \} }$ denotes the indicator function of the  event $\{ \cdot \}$.
  Furthermore, it is easy to see that $\E{ || A(s,t) || } = 2r \bar{v}_s (t-s)$.

  Second, we show that the number of sensors hitting the intruder during disjoint time intervals are independent. 
  This is simply done by observing that if $[s_1,t_1] \cap [s_2,t_2]=\emptyset$, each sensor is either selected in $\Phi(s_1,t_1)$ or in $\Phi(s_2,t_2)$ or not selected at all. Therefore, $\Phi(s_1,t_1)$ and $\Phi(s_2,t_2)$ are two independent processes.

  Combining the two properties, we conclude that the sequence of times at which the intruder gets hit is a Poisson process.

\hfill $\Box$

Compared to the case of stationary sensors where an undetected intruder
always remains undetected, the probability that the intruder is not detected 
in a mobile sensor network decreases exponentially over time,
\begin{eqnarray*}
P(X\geq t) = e^{-2\lambda r v_s t}.
\end{eqnarray*}
where $X$ represents the detection time of the intruder.

The expected detection  time of a randomly located intruder is
$E[X] = \frac{1}{2\lambda r v_s}$, which is inversely proportional to the
density of the sensors ($\lambda$), the sensing range of each sensor
($r$), and the speed of sensors ($v_s$).  Note that the
expected intruder detection time  
is independent of the sensor movement direction distribution density function,
$f_{\Theta}^{s}(\theta)$. Therefore, in order to quickly detect a stationary
intruder, one can add more sensors, use sensors with larger
sensing ranges, or increase the speed of the mobile sensors.

To guarantee that the expected time to detect a randomly located stationary 
intruder be smaller than a specific value $T_0$, we have 
\begin{eqnarray*}
\frac{1}{2\lambda r v_s} \leq T_0
\end{eqnarray*}
or equivalently,
\begin{eqnarray*}
\lambda  v_s \geq \frac{1}{2 r T_0}.
\end{eqnarray*}

If the sensing range of each sensor is fixed, the above formula presents the 
tradeoff between sensor density and sensor mobility to ensure given expected 
intruder detection time requirement. The product of the sensor density and 
sensor speed should be larger than a constant. Therefore, sensor mobility 
can be exploited to compensate for the lack of sensors, and vice versa.

In the proof of Theorem \ref{theo:msc}, we pointed out
that a location alternates between being covered and not being covered, 
and then derived the fraction of time that a point is covered. While
the time average characterization shows, to a certain extent, how well a 
point is covered, it does not reveal the duration of the time that a 
point is covered and uncovered. The time scales of such time durations 
are also very important for network planning; they present the time 
granularity of the intrusion detection capability that a mobile sensor 
network can provide. Theorem \ref{theo:sdt} now allows us to characterize 
the time durations of a point being covered and not being covered.

{\bf
\begin{corollary}
Consider a random sensor network $B(\lambda, r)$ at time $t=0$,
with sensors moving according to the straight-line random mobility model. 
A point alternates between being covered and not being covered. Denote 
the time duration that a point is covered as $T_c$, and 
the time duration that a point is not covered as $T_n$, we have
\begin{eqnarray}
T_n &\sim& \exp(2\lambda r v_s)\\ 
E[T_c] &=& \frac{e^{\lambda \pi r^2}-1}{2\lambda r v_s } .\label{eq:Tc}
\end{eqnarray}
\end{corollary}
}

{\bf Proof.}
In the proof of Theorem \ref{theo:sdt}, we know that the sequence of times 
at which a new sensor hits a given point forms a Poisson process of intensity 
$2\lambda r v_s$. After each sensor hits the point, it immediately covers the 
point until it moves out of range. There is no constraint on the number of 
sensors that covere the point. Therefore, the covered/uncovered sequence 
experienced by the point can be seen as a $M/G/\infty$ queuing process, where 
the service time of an sensor is the time duration that the sensor covers the 
point before moving out of range. The idle periods of $M/G/\infty$ queue 
corresponds to the time duration that the point is not covered. It is known that 
idle periods in such queues have exponentially distributed durations. Therefore, 
we have $T_n \sim \exp(2\lambda r v_s)$.

Since a point alternates between being 
covered and not being covered, the fraction of time a point is covered is 
\begin{eqnarray*}
f_t = \frac{E[T_c]}{E[T_c]+E[T_n]} = 1-e^{-\lambda \pi r^2}.
\end{eqnarray*}
The last equality in the above equation is given in (\ref{eq:ft}). Solving 
for $E[T_c]$, we obtain (\ref{eq:Tc}). 

Let $T$ denote the period of a point being covered and not being covered, i.e.,
$T=T_c+T_n$. The expected value of the period is 
\begin{eqnarray*}
E[T] = E[T_c] + E[T_n] = e^{\lambda \pi r^2}/2\lambda r v_s.
\end{eqnarray*}

\hfill $\Box$

\begin{figure} 
  \begin{center}
    \includegraphics[width=2in]{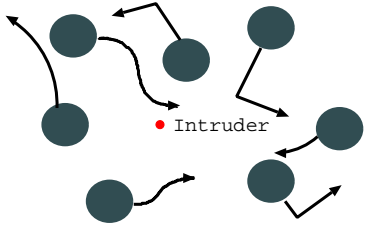}
    \caption{Mobile sensor network with sensors moving along arbitrary curves.}
    \label{fig:random_optimal}
  \end{center}
\end{figure}

Above we obtain the detection time of a stationary intruder when sensors all
move in straight lines. In practice, mobile sensors do not always move in 
straight lines; they may make turns and move in different curves, as depicted 
in Figure \ref{fig:random_optimal}. Next, we establish the optimal sensor 
moving strategy to minimize the detection time of a stationary intruder.

{\bf
\begin{theorem} \label{theo:optimal}
Consider a sensor network $B(\lambda, r)$ at time $t=0$, with sensors 
moving according to the random mobility model at a fixed speed $v_s$. 
The detection time of a randomly located stationary intruder, $X$, is 
minimized in probability if sensors all move in straight lines. 
\end{theorem}
}
{\bf Proof:}
  From Equation (\ref{eq:intensity}), we know that the number of sensors detecting the intruder during the interval $[0,t]$ is Poisson distributed with mean $\E{ || A(0,t) || }$. 
  Thus we have
  $$
  \pr{X\leq t} = \pr{\text{card}(\Phi(0,t)) \geq 1} = 1-\exp(-\E{ || A(0,t) || }),
  $$
  which is a increasing function of $\E{ || A(0,t) || }$.
  As $\E{ || A(0,t) || }$ is maximized when sensors move along straight lines, the probability of detecting the intruder is also maximized.

\hfill $\Box$

Similar to the arguments on the optimal strategies for area coverage 
in Section \ref{sec:areaCoverage}, straight line movement is not the 
only optimal strategy that minimizes the detection time. There is a 
family of moving patterns that can minimize the detection time, where
straight line movement is one of them.

In the above analysis, we have assumed that an intruder is immediately 
detected when it is hit by the perimeter of a sensor, regardless 
of the time duration ($t_s$) it stays in the sensing range of the sensor. 
In many intrusion detection applications, for example, radiation, chemical,
and biology threats, due to the probabilistic nature of the phenomenon and
the sensing mechanisms, an intruder will not be immediately detected once
it enters the sensing range of a sensor. Instead, it will take a certain 
amount of time to detect the intruder. If the sensing time 
is too short, an intruder may escape undetected. To account for this sensing 
time requirement, we define $t_d$ to be the minimum sensing time in order for a 
sensor to detect an intruder. Obviously, it is only interesting when $0 \leq t_d \leq 2r/v_s$. 
Otherwise, the sensing time of an intruder by a sensor will be smaller 
than the minimum requirement $t_d$, and the intruder will never be detected. 
In order to yield closed-form results and provide insights, we will consider
the straight-line random mobility model.

{\bf
\begin{theorem} \label{theo:dtwmstr}
Consider a sensor network $B(\lambda, r)$ at time $t=0$, with sensors moving 
according to the straight-line random mobility model at a fixed speed $v_s$. 
An intruder is detected iff the sensing time $t_s$ is at least $t_d$, 
i.e., $t_s\geq t_d$. Let $Y$ be the detection time of a randomly located 
stationary intruder initially not located in the sensing area of any sensor, 
we have
\begin{eqnarray} \label{eq:sdtmd}
Y = t_d + T
\end{eqnarray} 
where 
\begin{eqnarray}
T & \sim & \exp(2\lambda r_{\rm eff} v_s) \label{eq:T}\\
r_{\rm eff} & = & \sqrt{r^2-\frac{v_s^2t_d^2}{4}} \label{eq:effectiveRadius} .
\end{eqnarray}
\end{theorem}
}

\begin{figure}
 \begin{center}
  \includegraphics[width=3in]{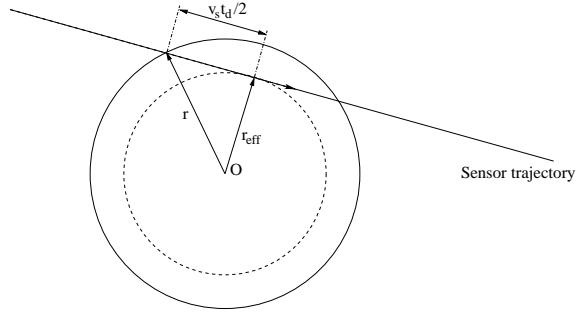}
  \caption{Effective radius of a mobile sensor.}
  \label{fig:reff}
 \end{center}
\end{figure}

{\bf Proof:}
  We assume without loss of generality that the intruder is located at the origin.
  We observe first that a sensor covers the intruder for a time longer than $t_d$ if and only if the distance between its trajectory and the origin is less than $r_\text{eff}$ (see Figure \ref{fig:reff}).
  We call such sensors \emph{valid} sensors.

  Similarly as in Theorem 1, we define a thinned Poisson process $\Phi_\text{eff}(0,t)$ by selecting the sensors that will detect the intruder during the interval $[0,t]$. 
  To do so, we define the \emph{effective} covered area $A_\text{eff}(0,t)$ of a sensor as the area covered by the disk of radius $r_\text{eff}$ centered on it.
  Then, the probability that a sensor initially located at $\bvec{x}$ detects the intruder during the interval $[0,t]$ is $\pr{-\bvec{x} \in A_\text{eff}(0,t)}$.
  By (\ref{eq:intensity}) we find that the expected number of points in $\Phi_\text{eff}(0,t)$ is $\lambda \E{ || A_\text{eff}(0,t) || } = 2\lambda r_\text{eff} v_s$. 
  Denoting by $T$ the time before a valid sensor covers the intruder, we get
  $$
  \pr{T\leq t} = \pr{\text{card}(\Phi_\text{eff}(0,t)) \geq 1} = 1-\exp(-2\lambda r_\text{eff} v_s).
  $$
  Then, the intruder is finally detected by the system after a time $T+t_d$.

\hfill $\Box$

In (\ref{eq:sdtmd}), the detection time has two terms, namely, a constant term 
$t_d$ and an exponentially distributed random variable with mean 
$E[T]= 1/(2\lambda r_{\rm eff} v_s)$. The first term $t_d$ is a direct consequence 
of the minimum sensing time requirement. After the perimeter of a sensor hits an intruder, 
it takes a minimum sensing time of $t_d$ to detect the intruder, and hence the constant 
delay. By Theorem \ref{theo:sdt}, the second term corresponds to the detection time in 
the case where there is no minimum sensing 
time requirement but sensors have a reduced sensing radius of $r_{\rm eff}$. This 
is again a consequence of the minimum sensing time requirement and the effect is 
illustrated in Figure \ref{fig:reff}. An intruder will only be detected 
by a mobile sensor if the trajectory of the sensor falls within $r_{\rm eff}$ from 
the intruder. The above two effects of minimum sensing time requirement result in an 
increased expected detection time compared to the case without minimum sensing time 
requirement. Since $t_d > 0$ and $r_{\rm eff} < r$, we have
\begin{eqnarray*}
E[Y] =  t_d + 1/(2\lambda r_{\rm eff} v_s) >  1/(2\lambda r v_s) = E[X].
\end{eqnarray*}

Sensor speed has two opposite effects on an intruder's detection time. 
\begin{itemize}
\item On one hand, as sensors move faster, uncovered areas will be covered 
more quickly and this tends to speed up the detection of intruders. 

\item On the other hand, the effective sensing radius $r_{\rm eff}$ decreases 
as sensors increase their speed due to the sensing time requirement, 
making intruders less likely to be detected. 
\end{itemize}

In the following, we
present the optimal sensor speed that minimizes the expected detection 
time. Excess mobility will be harmful when the sensor speed is larger than 
the optimal value.

{\bf
\begin{theorem}
Under the scenario in Theorem \ref{theo:dtwmstr}, the optimal sensor 
speed minimizing the expected detection time of a randomly located 
intruder is 
\begin{eqnarray}
v_s^* = \sqrt{2}r/t_d.
\end{eqnarray}
\end{theorem}
}
{\bf Proof.}
Let $d Y / d v_s = 0$,  we have $v_s^{*} = \sqrt{2}r/t_d$, and 
the second order derivative $\frac{d^2 Y}{dv_s^2}|_{v_s^*} < 0 $. 
The corresponding minimum expected detection time is 
\begin{eqnarray*}
E[Y^*] = (1+2\lambda r^2)t_d / 2\lambda r^2.
\end{eqnarray*}
\hfill $\Box$

\section{Detection Time of Mobile Intruder}\label{sec:MobileDetectionTime}
In this section, we consider the detection time of a mobile intruder, which 
depends not only on the mobility behavior of the sensors but also on the 
movement of the intruder itself. Intruders can adopt a wide variety of 
movement patterns. In this work, we will not consider specific intruder 
movement patterns. Rather, we approach the problem from a game theoretic 
standpoint and study the optimal mobility strategies of the intruders
and sensors. 

From Theorem \ref{theo:optimal}, the detection time of a stationary 
sensor intruder is minimized when sensors all move in straight lines. 
This result can be easily extended to a mobile intruder using similar
arguments in the reference framework where the intruder is stationary. 
From the perspective of an intruder, since it only knows the 
mobility strategy of the sensors (sensor direction distribution density 
function) and does not know the locations and directions of the sensors, 
changing direction and speed will not help prolong its detection time. 
In the following, we will only consider the case where sensors and 
intruders move in straight lines. 

Given the mobility model of the sensors, $f_{\Theta}(\theta)$, an
intruder chooses the mobility strategy that maximizes its expected
detection time. More specifically, an intruder chooses its speed 
$v_t \in [0, v_t^{\max})$  and direction $\theta_t \in [0, 2\pi)$ 
so as to maximize the expected detection time. The expected detection
time is a function of the sensor direction distribution density, intruder 
speed, and intruder moving direction. Denote the resulting expected detection
time as $\max_{v_t, \theta_t} E[X(f_{\Theta}(\theta), \theta_t,
  v_t)]$; the sensors then choose the mobility strategy (over all
possible direction distributions) that minimizes the maximum expected
 detection time. This can be viewed as a zero-sum 
minimax game between the 
collection of mobile sensors and the intruder, where the payoffs for the 
mobile sensors and intruder are $-E[X(f_{\Theta}, \theta_t, v_t)]$ and
$E[X(f_{\Theta}, \theta_t, v_t)]$, respectively.

To find the optimal mobility strategies for mobile
sensors and the intruder, we consider the following minimax optimization 
problem:  
\begin{eqnarray} \label{eq:minimax}
\min_{f_{\Theta}} \max_{\theta_t, v_t}
E[X(f_{\Theta}, \theta_t, v_t)].
\end{eqnarray}

To solve the minimax optimization problem, we first characterize
the detection time of an intruder moving at a constant speed in a
particular direction. 

{\bf
\begin{theorem}\label{theo:mdt}
Consider a sensor network $B(\lambda, r)$ at time $t=0$,
with sensors moving according to the stright-line random mobility model 
at a fixed speed $v_s$. Let $X$ be the detection time of an intruder 
moving at speed $v_t$ along direction $\theta_t$. Denote 
\begin{eqnarray*}
\begin{array}{lll}
c = v_t / v_s, \;\; \hat c  =  1+c \\ 
w(u)  =  \sqrt{1-\frac{4c}{\hat c ^2}\cos^2\frac{u}{2}} \\
\overline{v}_s = v_s \hat c \int_0^{2 \pi}
w(\theta-\theta_t) f_\Theta (\theta) d\theta . \\
\end{array}
\end{eqnarray*}
We have
\begin{eqnarray}\label{eq:mobile_vBar}
X\sim \exp(2\lambda r \overline{v}_s).
\end{eqnarray}
\end{theorem}
}

{\bf Proof:}
  To prove this theorem, we put ourselves in the frame of reference of the intruder and look at the speeds of the sensors. Thus, if a 
sensor has an absolute speed vector $\bvec{v}_s$, its speed vector in the new frame of reference is simply $\bvec{v}_s-\bvec{v}_t$, 
where $\bvec{v}_t$ denotes the intruder's absolute speed vector.  Let  $\theta_s$ denote the direction of $\bvec{v}_s$ 
and  $\theta_t$ the direction of $\bvec{v}_t$.

 In the new frame of reference, the intruder is static. Denote $c = v_t / v_s$ , $\hat c  =  1+c$, and 
$w(u)  =  \sqrt{1-\frac{4c}{\hat c ^2}\cos^2\frac{u}{2}}$. Using the Law of Cosines, the relative speed of the sensor can be computed as 
\begin{eqnarray*}
 || \bvec{v}_s-\bvec{v}_t || & = &  \sqrt{ v_s^2 + v_t^2 - 2v_s v_t cos (\theta_s-\theta_t)} \\
& = & v_s \hat c w(\theta_s-\theta_t)
\end{eqnarray*}

We know from Equation (\ref{eq:intensity}) that 
  $$
  \pr{X\leq t} = \pr{\text{card}(\Phi(0,t))\geq 1} = 1-\exp(-\lambda \E{ || A(0,t) || }).
  $$ 
  Therefore, if $\E{ || A(0,t) || }$ is a linear function of $t$, then $X$ is exponentially distributed.
  We get 
  $$
  || A(0,t) || = 2r || \bvec{v}_s-\bvec{v}_t || t =  2rt w(\theta_s-\theta_t),
  $$
  so that 
  \begin{eqnarray*}
  \E{|| A(0,t) ||} & = & 2rt \int_0^{2\pi} w(\theta-\theta_t)  f_\Theta(\theta) d\theta \\
 & = & 2rt \overline{v}_s
  \end{eqnarray*}
  where $\overline{v}_s = v_s \hat c \int_0^{2 \pi} w(\theta-\theta_t) f_\Theta(\theta) d\theta$, 
which can be viewed as the average effective sensor speed in the reference framework where the intruder 
is stationary. Therefore, the detection time is exponentially distributed with rate 
$2\lambda r \overline{v}_s$.

\hfill $\Box$

From Theorems \ref{theo:sdt} and  \ref{theo:mdt}, it can be
noted that the detection times of both stationary and mobile intruders
follow exponential distributions, and that the parameters are
of the same form, except that the sensor speed  is now
replaced by the effective sensor speed  for
the mobile intruder case.

Assuming that the sensor density and sensing range are fixed, since the
intruder detection time follows an exponential distribution with mean
 $1/(2\lambda r \overline{v}_s)$, maximizing the expected
detection time corresponds to minimizing the effective sensor speed
$\overline{v}_s$.  In the following, we derive the optimal intruder
mobility strategies for two special sensor mobility models.

\noindent {\bf Sensors move in the same direction $\theta_s$:
  $f_\Theta(\theta) = \delta(\theta-\theta_s)$.} 

Using the fundamental property of the delta function
$\int_{-\infty}^{\infty}f(x)\delta(x-a)dx=f(a)$, we have 

\begin{eqnarray*}
\overline{v}_s & = & v_s \hat c \int_0^{2 \pi} 
w(\theta-\theta_t) \delta(\theta-\theta_s)d\theta \\
& = &  v_s\hat c w(\theta_s-\theta_t).
\end{eqnarray*}

We need to choose a proper $\theta_t$ and $v_t$ that minimizes the above 
effective sensor speed $\overline{v}_s$. First, it is easy to see that we 
require $\theta_t = \theta_s$. Now, we have
\begin{eqnarray*}
\overline{v}_s =  v_s\hat c \sqrt{1-\frac{4c}{\hat c ^2}} = |v_t-v_s|
\end{eqnarray*}
and $\overline{v}_s$ is minimized when
\begin{eqnarray*}
v_t & = & \left\{ \begin{array}{ll} 
v_s & \mbox{ if  $v_t^{\max} \geq v_s$} \\
v_t^{\max} & \mbox{ otherwise.} 
\end{array} \right.
\end{eqnarray*}
 
The above results show, quite intuitively, that the intruder should move
in the same direction as the sensors at a speed closest matching the
sensor speed. If the maximum intruder speed is larger than the sensor
speed, the intruder will not be detected since it chooses to move at the
same speed and in the same direction as the sensors. In this case, the
detection time is infinity. Otherwise, if the maximum intruder speed is
smaller than the sensor speed, the intruder should move at the maximum
speed in the same direction of the sensors.  The expected detection time is  
$\frac{1}{2\lambda r (v_s-v_t^{\max})}$. 

\noindent {\bf Sensors move in uniformly random directions: $f_{\Theta} (\theta) = \frac{1}{2 \pi}$. }

\begin{figure} 
  \begin{center}
    \includegraphics[width=3in]{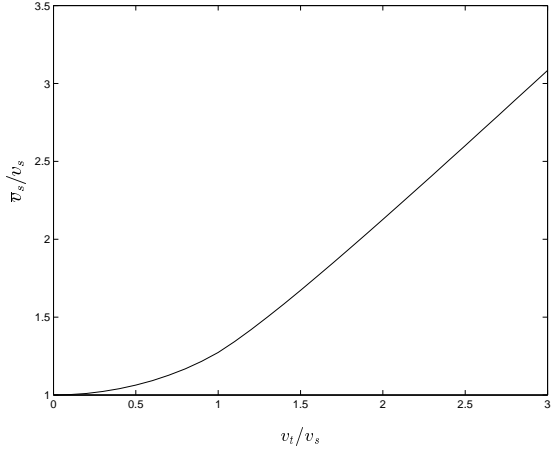}
    \caption{Normalized effective relative sensor speed
    $\overline{v}_s / v_s$ as a function of $c=v_t /v_s$ }
    \label{fig:vBar}
  \end{center}
\end{figure}

Figure \ref{fig:vBar} plots the normalized effective sensor speed
$\overline{v}_s / v_s$ as a function of $c=v_t/v_s$, the ratio of the 
intruder speed to the sensor speed. The effective sensor speed is an
increasing function of $c$, and is minimized when $c=0$, or $v_t=0$.
Therefore, if each sensor uniformly chooses its moving 
direction from 0 to $2\pi$, the maximum expected detection time is
achieved when the intruder does not move. The corresponding expected
detection time is $\frac{1}{2\lambda r v_s}$. The optimal intruder
mobility strategy in this case can be intuitively explained as
follows. Since sensors move in all directions with equal probability, 
the movement of the intruder in any direction will result in a larger
relative speed and thus a smaller first hit time in that particular
direction. Consequently, the minimum of the first hit times in all
directions (detection time) will become smaller.

We now present the solution to the minimax game between the collection
of mobile sensors and the intruder in the following theorem. 

{\bf
\begin{theorem} \label{theo:sos}
Consider a sensor network $B(\lambda, r)$ at time $t=0$, with sensors moving 
according to the random mobility model at a fixed speed $v_s$. For the game between 
the collection of mobile sensors and the intruder, the optimal sensor strategy is 
for each sensor to choose a direction according to a uniform distribution, i.e., 
$f_\Theta(\theta)  = \frac{1}{2 \pi}$. The optimal mobility strategy of the intruder 
is to stay stationary. This solution constitutes a Nash equilibrium of the game.
\end{theorem}
}

{\bf Proof.}
For sensors, minimizing intruder detection time is equivalent to
maximizing the effective sensor speed after an intruder selects the 
optimal speed and direction.  
We first prove for any given intruder speed $v_t$, that among all possible
sensor direction distributions, the minimum effective
sensor speed resulted from the optimal intruder direction choice, 
$\min_{\theta_t} \overline{v}_s$, is maximized when sensors choose 
directions according to a uniform distribution. The formal statement
is described as follows. 

Denote the uniform distribution density as $f_{\Theta}^{\rm uniform} =1/2\pi$. 
From Theorem \ref{theo:mdt}, the effective sensor speed is a function
of sensor direction distribution density, intruder speed and direction,
$\overline{v}_s (f_{\Theta}(\theta), \theta_t, v_t) =
\int_{0}^{2\pi}w(\theta-\theta_t)f_{\Theta}(\theta) d\theta$. 

We will prove that 
\begin{equation}
\label{optimization}
\min_{\theta_t,v_t}\overline{\nu}_s(f_{\Theta}(\theta), \theta_t, v_t)\leq
\min_{\theta_t,v_t}\overline{\nu}_s(f^{\rm uniform}_\Theta, \theta_t, v_t)
\end{equation}
for all $f_{\Theta}(\theta)$.

First, let us consider the right-hand side of (\ref{optimization}).
We have
\begin{eqnarray*}
\overline{\nu}_s(f^{\rm uniform}_\Theta, \theta_t, v_t)&=&
\frac{1}{2\pi}\int_0^{2\pi} w(\theta-\theta_t)\,d\theta
\nonumber\\
&=&
\frac{1}{2\pi}\int_{-\theta_t}^{2\pi-\theta_t} w(u)\, du
\nonumber\\
&=&
\frac{1}{2\pi}\int_0^{2\pi} w(u)\, du
\label{independent-theta-t}
\end{eqnarray*}
for all $\theta_t$, since the mapping $u\to w(u)$ is periodic 
with period $2\pi$.
This shows that 
\begin{equation}
\label{simplify-optimization-pb}
\min_{\theta_t}\overline{\nu}_s(f^{\rm uniform}_\Theta, \theta_t, v_t)
= \frac{1}{2\pi}\int_0^{2\pi} w(u)\, du.
\end{equation}
We now come back to the proof of (\ref{optimization}).
We have
\begin{eqnarray}
&&\min_{\theta_t}\overline{\nu}_s(f_{\Theta}(\theta), \theta_t, v_t)
\nonumber\\
&\leq &
\frac{1}{2\pi}\int_0^{2\pi} \overline{\nu}_s(f_{\Theta}(\theta), \theta_t, v_t)\, d\theta_t
\nonumber\\
&=&
\frac{1}{2\pi}\int_0^{2\pi}\int_0^{2\pi} w(\theta-\theta_t) f_{\Theta}(\theta) \,d\theta d\theta_t
\nonumber\\
&=&
\frac{1}{2\pi}\int_0^{2\pi} f_\Theta(\theta)
\left(\int_{\theta-2\pi}^{\theta} w(u)\, du\right)  d\theta
\nonumber\\
&=& 
\frac{1}{2\pi} \left(\int_0^{2\pi} f_\Theta(\theta)\,d\theta\right)
\left(\int_0^{2\pi} w(u)\, du\right)
\nonumber\\
&=&
\frac{1}{2\pi}\int_0^{2\pi} w(u)\,du
\nonumber\\
&=&\min_{\theta_t}\overline{\nu}_s(f^{\rm uniform}_\Theta, \theta_t, v_t)
\label{eqn1}
\end{eqnarray}
where the last three equalities follow from the fact that 
$w(u)$ is periodic with period $2\pi$, from the fact
that $f_\Theta(\theta)$ is a probability density function on $[0,2\pi]$,
and from (\ref{simplify-optimization-pb}), respectively.

The proof of (\ref{optimization}) is concluded by taking first the minimum over
$v_t$ in the left-hand side of (\ref{eqn1}), then by taking
the minimum over $v_t$ in the right-hand side of (\ref{eqn1}).

It follows  that when sensors choose directions according to a uniform 
distribution, the optimal intruder mobility strategy is to stay stationary,
$v_t=0$ (since $\overline{v}_s(f^{\rm uniform}_{\Theta},\theta_t,v_t)$  
is maximized when $c=0$ (and equals to $1$), i.e. when $v_t=0$), 
and $\theta_t$ is irrelevant in this case.

Based on the previous discussions on different mobility strategies of 
sensors and intruders, under the derived optimal mobility strategies,
neither side can improve the payoff by changing the strategy unilaterally.
Specifically, when sensors choose their direction uniformly at random, 
the movement of the intruder in any direction will result in a larger
relative speed and thus a smaller first hit time in that particular
direction. Consequently, the minimum of the first hit times in all
directions (detection time) will become smaller. When the intruder stays
stationary, the dection time will not improve if sensors choose a 
different distribution for the moving direction. Therefore, the solution
constitutes a Nash equilibrium of the game.
 
\hfill $\Box$

This result suggests that in order to minimize the expected detection
time of an intruder, sensors should choose their directions uniformly
at random between $[0, 2\pi)$. The corresponding optimal mobility strategy 
of the intruder is to stay stationary.

The uniformly random sensor movement represents a mixed strategy which is a 
Nash equilibrium of the game between mobile sensors and intruders. If sensors 
choose to move in any fixed direction (pure strategy), it can be exploited by 
an intruder by moving in the same direction as sensors to maximize its detection 
time. The optimal sensor strategy is to choose a mixture of available pure 
strategies (move in a fixed direction between $[0, 2\pi)$). The proportion 
of the mix should be such that the intruder cannot exploit the choice by 
pursuing any particular pure strategy (move in the same direction as 
sensors), resulting in a uniformly random distribution for sensor's movement. 
When sensors and intruders follow their respective optimal strategies, 
neither side can achieve better performance by deviating from this behavior.

\newpage
\section{Summary} \label{sec:conclusions}

In this paper, we study the dynamic aspects of the coverage of a mobile 
sensor network resulting from the continuous movement of sensors. 
Specifically, we studied the coverage measures related to the area 
coverage and intrusion detection capability of a mobile sensor network. 

For the random initial deployment and the random sensor mobility model 
under consideration, we showed that while the area coverage at any given 
time instants remains unchanged, more area will be covered at least once
during a time interval. This is important for applications that do not 
require or cannot afford simultaneous coverage of all locations but want
to cover the deployed region within a certain time interval. The cost is 
that a location is only covered part of the time, alternating between 
covered and not covered. To this end, we characterized the durations and 
fraction of time that a location is covered and not covered. 

As sensors move around, intruders that will never be detected in a stationary 
sensor network can be detected by moving sensors. We characterized the 
detection time of a randomly located stationary intruder. The results 
suggest that sensor mobility can be exploited to effectively reduce the 
detection time of an intruder when the number of sensors is limited. We 
further considered a more realistic sensing model where a minimum sensing 
time is required to detect an intruder. We find that there is an 
optimal sensor speed that minimizes the expected detection time. Beyond the
optimal speed, excess mobility will be harmful to the intrusion detection 
performance. Moreover, we discussed the optimal mobility strategies that 
maximize the area coverage during a time interval and minimize the detection 
time of intruders.

For mobile intruders, the intruder detection time depends on the mobility 
strategies of the sensors as well as the intruders. We took a game theoretic 
approach and obtained the optimal mobility strategy for sensors and intruders.
We showed that the optimal sensor mobility strategy is that each sensor 
chooses its direction uniformly at random in all directions. By 
maximizing the entropy of the sensor direction distribution, the amount of 
prior information on sensor mobility strategy revealed to an intruder is 
minimized. The corresponding intruder mobility strategy is to stay stationary 
in order to maximize its detection time. This solution represents a Nash 
equilibrium of the game between mobile sensors and intruders. Neither side 
can achieve better performance by deviating from their respective optimal
strategies.

\singlespacing
\bibliographystyle{IEEEtran.bst}
\bibliography{ref}

\end{document}